\documentclass[a4paper,aps,twocolumn]{revtex4}
\usepackage{amsmath}
\begin{document}
\title{\bf{Torsion, Magnetic Monopoles and Faraday's Law  via a Variational Principle}}
\author{Philip D. Mannheim}
\affiliation{Department of Physics, University of Connecticut, Storrs, CT 06269, USA.
email: philip.mannheim@uconn.edu}
\date{June 25, 2014}
\begin{abstract}
Even though Faraday's Law  is a dynamical law that describes how changing $\bf{E}$ and $\bf {B}$  fields influence each other, by introducing a vector potential $A_{\mu}$ according to $F_{\mu\nu}=\partial_{\mu}A_{\nu}-\partial_{\nu}A_{\mu}$ Faraday's Law is satisfied kinematically, with the relation $(-g)^{-1/2}\epsilon^{\mu\nu\sigma\tau}\nabla_{\nu}F_{\sigma\tau}=0$  holding on every path in a variational procedure or path integral. In a space with torsion $Q_{\alpha\beta\gamma}$  the axial vector $S^{\mu}=(-g)^{1/2}\epsilon^{\mu\alpha\beta\gamma}Q_{\alpha\beta\gamma}$ serves as a chiral analog of $A_{\mu}$, and via variation with respect to $S_{\mu}$ one can derive Faraday's Law dynamically as a stationarity condition. With $S_{\mu}$ serving as an axial potential one is able to introduce magnetic monopoles without $S_{\mu}$ needing to be singular or have a non-trivial topology.  Our analysis permits torsion and magnetic monopoles to be intrinsically Grassmann, which could explain why they have never been detected. Our procedure permits us to both construct a Weyl geometry in which $A_{\mu}$ is metricated and then convert it into a standard Riemannian geometry.

\end{abstract}
\maketitle

\section{Introduction}
The Maxwell equations of electromagnetism in flat space break up into two groups, the Maxwell-Ampere and Electric Gauss Laws 
\begin{eqnarray} 
\boldsymbol{\nabla} \times \boldsymbol{B}-\frac{\partial\boldsymbol{E}}{\partial t}=\boldsymbol{J}_e,\qquad \boldsymbol{\nabla}\cdot\boldsymbol{E}=\rho_e,
\label{1}
\end{eqnarray}
and the  Faraday and Magnetic Gauss Laws 
\begin{eqnarray}
 \boldsymbol{\nabla}\times \boldsymbol{E}+\frac{\partial\boldsymbol{B}}{\partial t}=0,\qquad \boldsymbol{\nabla}\cdot\boldsymbol{B}=0.
\label{2}
\end{eqnarray}
Since can one derive second-order wave equations for the propagation of the $\bf{E}$ and $\bf{B}$ field strengths in a source-free region only when all of the above equations are taken in conjunction, as such all of these equations should be regarded as being on an equal dynamical footing. And if they are to be on an equal dynamical footing, then each one of these equations should, like all dynamical equations, be derivable via stationary variation of an action.

However, the standard treatment of electrodynamics  is not formulated in this way,  as it does not in fact derive all of these equations from a variational procedure. Rather, in order to develop the variational procedure that it does use, it relies on the fact that the Faraday-Magnetic Gauss equations immediately admit of an exact solution
\begin{eqnarray}
\boldsymbol{E}=-\frac{\partial{\boldsymbol{A}}}{\partial t}-\boldsymbol{\nabla}\phi,\qquad  \boldsymbol{B}=\boldsymbol{\nabla}\times\boldsymbol{A},
\label{3}
\end{eqnarray}
a solution that is unique up to gauge transformations of the form $\boldsymbol{A} \rightarrow \boldsymbol{A}+\boldsymbol{\nabla}\chi$, $\phi \rightarrow \phi -\partial \chi/\partial t$. The introduction of  $\bf{A}$ and $\phi$ serves two purposes. When inserted into Eq. (\ref{1}) they enable one to solve for  the $\bf{E}$ and $\bf{B}$ fields once  $\mathbf{J}_e$ and $\rho_e$ are specified. And in addition they allow one to develop a variational procedure.

To discus the variational procedure it is more convenient to first write the Maxwell equations covariantly in a curved space where they generalize to
\begin{eqnarray}
\nabla_{\nu}F^{\nu\mu}=J^{\mu},
\label{4}
\end{eqnarray}
\begin{eqnarray}
&&(-g)^{-1/2}\epsilon^{\mu\nu\sigma\tau}\nabla_{\nu}F_{\sigma\tau}
=0.
\label{5}
\end{eqnarray}
In Eq. (\ref{5}) the antisymmetric rank two tensor $F^{\mu\nu}$ is the field strength with components $F^{01}=-E_x$, $F^{12}=-B_z$ etc., and $J^{\mu}=(\rho_e,\mathbf{J}_e)$. Using 
$-(-g)^{-1}\epsilon_{\mu\nu\sigma\tau}\epsilon^{\mu\alpha\beta\gamma}=
\delta^{\alpha}_{\nu}\delta^{\beta}_{\sigma}\delta^{\gamma}_{\tau}+
\delta^{\alpha}_{\tau}\delta^{\beta}_{\nu}\delta^{\gamma}_{\sigma}+
\delta^{\alpha}_{\sigma}\delta^{\beta}_{\tau}\delta^{\gamma}_{\nu}-
\delta^{\alpha}_{\nu}\delta^{\beta}_{\tau}\delta^{\gamma}_{\sigma}-
\delta^{\alpha}_{\sigma}\delta^{\beta}_{\nu}\delta^{\gamma}_{\tau}-
\delta^{\alpha}_{\tau}\delta^{\beta}_{\sigma}\delta^{\gamma}_{\nu}$, Eq. (\ref{5}) can also  be written in the form
\begin{eqnarray}
\nabla_{\nu}F_{\sigma\tau}+\nabla_{\tau}F_{\nu\sigma}+\nabla_{\sigma}F_{\tau\nu}=0.
\label{6}
\end{eqnarray}
For brevity we shall refer to Eq. (\ref{5})  as Faraday's Law even as it encompass Gauss' Law of Magnetism as well. With Eq. (\ref{5}) possessing an exact solution of the form $F_{\mu\nu}=\nabla_{\mu}A_{\nu}-\nabla_{\nu}A_{\mu}$, one introduces the Maxwell  action
\begin{eqnarray}
I_{\rm MAX}=\int d^4x(-g)^{1/2}\bigg{[} -\frac{1}{4}F_{\mu\nu}F^{\mu\nu}-A_{\mu}J^{\mu}\bigg{]},
\label{7}
\end{eqnarray}
with its stationary variation with respect to $A_{\mu}$ immediately leading to Eq. (\ref{4}).

Since this variation is a variation in which Eq. (\ref{5}) is  not varied, Eq. (\ref{5}) is required to hold on every variational path. Thus even though Faraday's Law is a dynamical equation, the variation that is done is a constrained one in which Faraday's Law is imposed on every variational path, even on those that are not stationary, with the quantum path integral $\int DA_{\mu}\exp(iI_{\rm MAX})$ then being  constrained this way as well. We shall thus seek to construct a variational procedure in which Faraday's Law is to only hold at the stationary minimum.

\section{Setting up the Variational Procedure}

If we do not want Faraday's Law to hold on non-stationary paths, we cannot set $F_{\mu\nu}=\nabla_{\mu}A_{\nu}-\nabla_{\nu}A_{\mu}$, since this would immediately cause $(-g)^{-1/2}\epsilon^{\mu\nu\sigma\tau}\nabla_{\nu}F_{\sigma\tau}$ to vanish \cite{footnote1}. If however, we wish to recover Faraday's Law at the stationary minimum, then with 8 equations being embodied in Eqs. (\ref{4}) and (\ref{5}), we need not one but two 4-vector potentials, one of course being the standard vector potential $A_{\mu}$ and the other needing to be some as yet to be identified axial vector $S_{\mu}$. Moreover, without regard to variational issues, we note that in the event of magnetic monopoles one would ordinarily (though not quite in fact as we show below) modify Eq. (\ref{5}) to  
\begin{eqnarray}
(-g)^{-1/2}\epsilon^{\mu\nu\sigma\tau}\nabla_{\nu}F_{\sigma\tau}=K^{\mu},
\label{8}
\end{eqnarray}
with there then being both vector and axial vector current sources, for a total of 8 components. In the same way as we couple $A_{\mu}$ to $J^{\mu}$ via $A_{\mu}J^{\mu}$ we should equally anticipate a coupling $S_{\mu}K^{\mu}$ in the axial current sector, a coupling that is parity conserving if $S_{\mu}$ is an axial vector.  The issue of constructing a variational principle for Faraday's Law is thus related to the coupling of electromagnetism to magnetic currents, and our objective will be to set up a variational principle with respect to $A_{\mu}$ and $S_{\mu}$ that would recover Eqs. (\ref{4}) and (\ref{8}) at the stationary minimum, with Eq. (\ref{5}) then following in the limit in which we could set the monopole current to zero.

Recalling the two-potential study \cite{Shanmugadhasan1952,Cabibbo1962} of the monopole problem \cite{footnote2}, it is very convenient to introduce 
\begin{eqnarray}
X^{\mu\nu}&=&\nabla^{\mu}A^{\nu}-\nabla^{\nu}A^{\mu}
\nonumber\\
&-&\frac{1}{2}(-g)^{-1/2}\epsilon^{\mu\nu\sigma\tau}(\nabla_{\sigma}S_{\tau}-\nabla_{\tau}S_{\sigma})
\label{9}
\end{eqnarray}
as a generalized $F^{\mu\nu}$. On setting $S^{\mu\nu}=\nabla^{\mu}S^{\nu}-\nabla^{\nu}S^{\mu}$, we can rewrite $X^{\mu\nu}$ in terms of $F^{\mu\nu}$ and the dual $\hat{S}^{\mu\nu}=(1/2)(-g)^{-1/2}\epsilon^{\mu\nu\sigma\tau}S_{\sigma\tau}$ of $S^{\mu\nu}$ according to:
\begin{eqnarray}
X^{\mu\nu}=F^{\mu\nu}-\hat{S}^{\mu\nu},~~~\hat{X}^{\mu\nu}=\hat{F}^{\mu\nu}+S^{\mu\nu}.
\label{10}
\end{eqnarray}
(If $\epsilon^{0123}=+1$, $\epsilon_{0123}=-1$.)
Given this $X^{\mu\nu}$, Eqs. (\ref{4}) and ({\ref{8}) are to be replaced by 
\begin{eqnarray}
\nabla_{\nu}X^{\nu\mu}&=&\nabla_{\nu}F^{\nu\mu}=J^{\mu},\qquad \nabla_{\nu}\hat{X}^{\nu\mu}=\nabla_{\nu}S^{\nu\mu}=K^{\mu},
\nonumber\\
\nabla_{\nu}\hat{F}^{\nu\mu}&=&0,\qquad \nabla_{\nu}\hat{S}^{\nu\mu}=0,
\label{11}
\end{eqnarray}
with it now being $\nabla_{\nu}\hat{X}^{\nu\mu}=K^{\mu}$ and not in fact $\nabla_{\nu}\hat{F}^{\nu\mu}=K^{\mu}$ that is to describe the monopole. If we introduce a second set of field strengths $S^{01}=-B_x^{\prime}$, $S^{12}=+E_z^{\prime}$, $\hat{S}^{01}=E_x^{\prime}$, $\hat{S}^{12}=B_z^{\prime}$, on setting $K^{\mu}=(\rho_m,-\mathbf{J}_m)$, we find that in flat space Eq. (\ref11}) breaks up into two sectors, namely  Eqs. (\ref{1}) and (\ref{2}) and the analog
\begin{eqnarray} 
\boldsymbol{\nabla} \times \boldsymbol{B}^{\prime}-\frac{\partial\boldsymbol{E}^{\prime}}{\partial t}=0,\qquad \boldsymbol{\nabla}\cdot\boldsymbol{E}^{\prime}=0,
\nonumber\\
 \boldsymbol{\nabla}\times \boldsymbol{E}^{\prime}+\frac{\partial\boldsymbol{B}^{\prime}}{\partial t}=\boldsymbol{J}_m,\qquad \boldsymbol{\nabla}\cdot\boldsymbol{B}^{\prime}=\rho_m.
\label{12}
\end{eqnarray}
Moreover, if we define $\bf{E}_{\rm TOT}=\bf{E}+\bf{E}^{\prime}$, $\bf{B}_{\rm TOT}=\bf{B}+\bf{B}^{\prime}$, we can combine Eqs. (\ref{1}), (\ref{2}), and (\ref{12}) into 
\begin{eqnarray} 
\boldsymbol{\nabla} \times \boldsymbol{B}_{\rm TOT}-\frac{\partial\boldsymbol{E}_{\rm TOT}}{\partial t}=\boldsymbol{J}_e,\qquad \boldsymbol{\nabla}\cdot\boldsymbol{E}_{\rm TOT}=\rho_e,
\nonumber\\
 \boldsymbol{\nabla}\times \boldsymbol{E}_{\rm TOT}+\frac{\partial\boldsymbol{B}_{\rm TOT}}{\partial t}=\boldsymbol{J}_m,\qquad \boldsymbol{\nabla}\cdot\boldsymbol{B}_{\rm TOT}=\rho_m.
\label{13}
\end{eqnarray}
Thus even if $\mathbf{J}_m$ and $\rho_m$ can be neglected, it is $\mathbf{E}_{\rm TOT}$ and $\mathbf{B}_{\rm TOT}$ that are measured in electromagnetic experiments.

On introducing the action 
\begin{eqnarray}
I=\int d^4x(-g)^{1/2}\bigg{[} -\frac{1}{4}X_{\mu\nu}X^{\mu\nu}-A_{\mu}J^{\mu}-S_{\mu}K^{\mu}\bigg{]},
\label{14}
\end{eqnarray}
we find that stationary variation with respect to $A_{\mu}$ and $S_{\mu}$ then immediately leads to Eq. (\ref{11}), just as we want. Moreover, up to surface terms this action decomposes into two sectors according to 
\begin{eqnarray}
I&=&\int d^4x(-g)^{1/2}\bigg{[} -\frac{1}{4}F_{\mu\nu}F^{\mu\nu}-A_{\mu}J^{\mu}
\nonumber\\
&-&\frac{1}{4}S_{\mu\nu}S^{\mu\nu}-S_{\mu}K^{\mu}\bigg{]}.
\label{15}
\end{eqnarray}
Thus with the introduction of a magnetic current sector we can formulate a variational principle for Faraday's Law and for theories that involve magnetic monopoles, and can do so without the use of singular potentials or non-trivial topologies \cite{footnote4}. However, we still need to ascribe a physical meaning to $S_{\mu}$, and to this end we turn to torsion. This will lead us directly to the action given in Eq. (\ref{15}), and suggest a rationale for why the $S_{\mu\nu}$ sector has escaped detection and why a purely $A_{\mu}$-based quantum electrodynamics works as well as it does.

\section{Torsion}

To construct covariant derivatives in a metric theory one introduces a connection $\Gamma^{\lambda}_{\phantom{\alpha}\mu\nu}$. For a torsionless Riemann space one uses the Levi-Civita and spin connections
\begin{eqnarray}
\Lambda^{\lambda}_{\phantom{\alpha}\mu\nu}&=&\frac{1}{2}g^{\lambda\alpha}(\partial_{\mu}g_{\nu\alpha} +\partial_{\nu}g_{\mu\alpha}-\partial_{\alpha}g_{\nu\mu})=\Lambda^{\lambda}_{\phantom{\alpha}\nu\mu},
\nonumber\\
-\omega_{\mu}^{ab}&=&V^b_{\nu}\partial_{\mu}V^{a\nu}+V^b_{\lambda}\Lambda^{\lambda}_{\phantom{\lambda}\nu\mu}V^{a\nu}=\omega_{\mu}^{ba},
\label{17}
\end{eqnarray}
to construct covariant derivatives such as $\nabla_{\mu}g^{\lambda\nu}=\partial_{\mu}g^{\lambda\nu}+\Lambda^{\lambda}_{\phantom{\alpha}\alpha\mu}g^{\alpha\nu}+\Lambda^{\nu}_{\phantom{\alpha}\alpha\mu}g^{\lambda\alpha}$ and $D_{\mu}V^{a\lambda}=\partial_{\mu}V^{a\lambda}+\Lambda^{\lambda}_{\phantom{\alpha}\nu\mu}V^{a \nu}+\omega_{\mu}^{ab}V^{\lambda}_{b}$ that transform as tensors under local translations and local Lorentz transformations. In Eq. (\ref{17}) we have introduced vierbeins $V^{a}_{\mu}$ that carry an index $a$ associated with a fixed special-relativistic reference system, with the metric being writable as $g_{\mu\nu}=\eta_{ab}V^{a}_{\mu}V^{b}_{\nu}$. The covariant derivatives of $g_{\mu\nu}$ and $V^{\mu a}$  constructed with $\Lambda^{\lambda}_{\phantom{\alpha}\mu\nu}$ obey the metricity conditions $\nabla_{\mu}g^{\lambda\nu}=0$, $D_{\mu}V^{a\lambda}=0$. If one generalizes $\Lambda^{\lambda}_{\phantom{\alpha}\mu\nu}$ to $\tilde{\Gamma}^{\lambda}_{\phantom{\alpha}\mu\nu}$ by adding a rank-3 tensor to it,  covariant derivatives constructed with $\tilde{\Gamma}^{\lambda}_{\phantom{\alpha}\mu\nu}$ will still transform as true tensors. However, they may not necessarily obey metricity conditions $\tilde{\nabla}_{\mu}g^{\lambda\nu}=0$, $\tilde{D}_{\mu}V^{a\lambda}=0$ with respect to $\tilde{\Gamma}^{\lambda}_{\phantom{\alpha}\mu\nu}$.

To extend the geometry to include torsion one takes the connection to no longer be symmetric on its two lower indices, and defines the Cartan torsion tensor $Q^{\lambda}_{\phantom{\alpha}\mu\nu}$  
\begin{eqnarray}
Q^{\lambda}_{\phantom{\alpha}\mu\nu}=\Gamma^{\lambda}_{\phantom{\alpha}\mu\nu}-\Gamma^{\lambda}_{\phantom{\alpha}\nu\mu}.
\label{21}
\end{eqnarray}
To implement metricity one defines a contorsion tensor 
\begin{eqnarray}
K^{\lambda}_{\phantom{\alpha}\mu\nu}=\frac{1}{2}g^{\lambda\alpha}(Q_{\mu\nu\alpha}+Q_{\nu\mu\alpha}-Q_{\alpha\nu\mu}),
\label{22}
\end{eqnarray}
and with $K^{\lambda}_{\phantom{\alpha}\mu\nu}$ one constructs connections of the form
\begin{eqnarray}
\tilde{\Gamma}^{\lambda}_{\phantom{\alpha}\mu\nu}&=&\Lambda^{\lambda}_{\phantom{\alpha}\mu\nu}+K^{\lambda}_{\phantom{\alpha}\mu\nu},
\nonumber\\
-\tilde{\omega}_{\mu}^{ab}&=&-\omega_{\mu}^{ab}+V^{b}_{\lambda}K^{\lambda}_{\phantom{\alpha}\nu\mu}V^{a \nu}=\tilde{\omega}_{\mu}^{ba}.
\label{23}
\end{eqnarray}

To couple spinors to gravity in a Riemannian space without torsion one uses the covariantized Dirac action $I _{\rm D}=(1/2)\int d^4x(-g)^{1/2}i\bar{\psi}\gamma^{a}V^{\mu}_a(\partial_{\mu}+\Sigma_{bc}\omega^{bc}_{\mu})\psi +H. c.$, where $\Sigma_{ab}=(1/8)(\gamma_a\gamma_b-\gamma_b\gamma_a)$. To generalize this action to include torsion one replaces $\omega^{bc}_{\mu}$ by $\tilde{\omega}^{bc}_{\mu}$ and obtains 
\begin{eqnarray}
\tilde{I}_{\rm D}=\frac{1}{2}\int d^4x(-g)^{1/2}i\bar{\psi}\gamma^{a}V^{\mu}_a(\partial_{\mu}+\Sigma_{bc}\tilde{\omega}^{bc}_{\mu})\psi+H. c.~~
\label{25}
\end{eqnarray}
Integration parts, use of properties of the Dirac gamma matrices, and introduction of a coupling to $A_{\mu}$ yields \cite{Shapiro2002}
\begin{eqnarray}
\tilde{I}_{\rm D}&=&\int d^4x(-g)^{1/2}i\bar{\psi}\gamma^{a}V^{\mu}_a(\partial_{\mu}+\Sigma_{bc}\omega^{bc}_{\mu}
\nonumber\\
&-&iA_{\mu}-i\gamma^5S_{\mu})\psi,
\label{26}
\end{eqnarray}
where 
\begin{eqnarray}
&&S^{\mu}=\frac{1}{8}(-g)^{-1/2}\epsilon^{\mu\alpha\beta\gamma}Q_{\alpha\beta\gamma},
\nonumber\\
&&-(-g)^{-1/2}\epsilon_{\mu\alpha\beta\gamma}S^{\mu}=\frac{1}{4}[Q_{\alpha\beta\gamma}+Q_{\gamma\alpha\beta}+Q_{\beta\gamma\alpha}].~~~
\label{27}
\end{eqnarray}
In the action $\tilde{I}_{\rm D}$  we note that even though the torsion is only antisymmetric on two of its indices, the only components of the torsion that appear in its torsion-dependent $S^{\mu}$ term are the four that constitute that part of the torsion that is antisymmetric on all three of its indices. As  well as being locally gauge invariant under $\psi\rightarrow e^{i\alpha(x)}\psi$, $A_{\mu}\rightarrow A_{\mu}+\partial_{\mu}\alpha(x)$,  $\tilde{I}_{\rm D}$ is also locally chiral invariant \cite{Shapiro2002} under $\psi\rightarrow e^{i\gamma^5\beta(x)}\psi$, $S_{\mu}\rightarrow S_{\mu}+\partial_{\mu}\beta(x)$. Additionally, as noted in \cite{Fabbri2014}, $\tilde{I}_{\rm D}$ is locally conformal invariant under $V^a_{\mu}(x)\rightarrow \Omega(x)V^a_{\mu}(x)$, $\psi(x)\rightarrow \Omega^{-3/2}(x)\psi(x)$ since, just like the vector potential $A_{\mu}$, the equally minimally coupled $S_{\mu}$ also has zero conformal weight \cite{footnote5}. The $\tilde{I}_{\rm D}$ action thus has a remarkably rich local invariance structure, as it is invariant under local translations, local Lorentz transformations, local gauge transformations, local axial gauge transformations, and local conformal transformations. 

With $S^{\mu}$ having a structure identical to the Faraday Law structure given in Eqs. (\ref{5}) and (\ref{6}), and with $S^{\mu}$ precisely being an axial 4-vector, $S^{\mu}$  is thus the natural quantity to act as the second potential  that appears in $X_{\mu\nu}$ \cite{footnote6}, and thus the natural axial vector needed to set up a variational procedure for Faraday's Law of electromagnetism \cite{footnote7}. However, in order to set up a variational procedure we will need to construct a kinetic energy term for it. To generate such a kinetic energy term we appeal to the Dirac action. Specifically,  we recall \cite{tHooft2010}, \cite{Shapiro2002} that when one does a path integration  $\int D\bar{\psi}D\psi\exp(i\tilde{I}_{\rm D})$ over the fermions (equivalent to a one fermion loop Feynman graph) one generates an effective action of the form \cite{footnote8}
\begin{eqnarray}
I_{\rm EFF}&=&\int d^4x(-g)^{1/2}C\bigg{[}\frac{1}{20}\left[R_{\mu\nu}R^{\mu\nu}-\frac{1}{3}(R^{\alpha}_{\phantom{\alpha}\alpha})^2\right]
\nonumber\\
&+&\frac{1}{3}F_{\mu\nu}F^{\mu\nu}+\frac{1}{3}S_{\mu\nu}S^{\mu\nu}\bigg{]},
\label{28}
\end{eqnarray}
where $C$ is a  log divergent constant and $R_{\mu\nu}$ is the standard (torsionless) Ricci tensor. The action $I_{\rm EFF}$ possesses all the  local symmetries possessed by $\tilde{I}_{\rm D}$, with the appearance of the $R_{\mu\nu}R^{\mu\nu}-(1/3)(R^{\alpha}_{\phantom{\alpha}\alpha})^2$ term being characteristic of a gravity theory that is locally conformal invariant (see e.g. \cite{Mannheim2006,Mannheim2012}). Also, we take note of the fact that path integration over the fermions has converted terms that are  linear in $A_{\mu}$ and $S_{\mu}$ in $\tilde{I}_{\rm D}$ into terms that are quadratic in $A_{\mu}$ and $S_{\mu}$ in $I_{\rm EFF}$. Comparing now with Eq. (\ref{15}), we see that the action $I_{\rm EFF}$ contains precisely the kinetic energy term we seek. Thus not only does torsion provide a natural origin for the second potential needed for $X_{\mu\nu}$, up to renormalization constants it also provides precisely the correct action whose variation, on adding appropriately coupled sources, leads to Eq. (\ref{11}) and a derivation of Faraday's Law via a variational principle. $S_{\mu}$ thus serves as an analog of the electromagnetic $A_{\mu}$, an  analog that is purely geometrical. 

Given the geometrical structure of $S_{\mu}$, we note that it is also possible to give $A_{\mu}$ an analogous such structure. Specifically, we recall that Weyl had suggested that one could metricate electromagnetism by introducing a $B_{\mu}$-dependent connection  for a real field $B_{\mu}$ of the form 
\begin{eqnarray}
W^{\lambda}_{\phantom{\alpha}\mu\nu}&=&-\frac{2}{3}g^{\lambda\alpha}(g_{\nu\alpha}B_{\mu} +g_{\mu\alpha}B_{\nu}-g_{\nu\mu}B_{\alpha})=W^{\lambda}_{\phantom{\alpha}\nu\mu},~~~~~~
\label{28a}
\end{eqnarray}
as written here with a convenient charge $2/3$ normalization. However, if we now use $\Lambda^{\lambda}_{\phantom{\alpha}\mu\nu}+K^{\lambda}_{\phantom{\alpha}\mu\nu}+W^{\lambda}_{\phantom{\alpha}\mu\nu}$ in the spin connection, as noted in \cite{Hayashi1977} the $B_{\mu}$ term drops out of the Dirac action identically, with Weyl's $B_{\mu}$ not coupling to the Dirac spinor at all. The reason for this is that the Weyl connection generates individual non-Hermitian terms of the generic form $i(\partial_{\mu}+B_{\mu})\psi$, and in the full Hermitian $\tilde{I}_{\rm D}$ such terms must cancel identically. However, given this, suppose we instead take $B_{\mu}$ to be anti-Hermitian and set $B_{\mu}=iA_{\mu}$ where $A_{\mu}$ is Hermitian. Now, not only is there now no cancellation, use of this anti-Hermitian connection is found to precisely lead to none other than the above $\tilde{I}_{\rm D}$ as given in Eq. (\ref{26}). Thus starting from the torsionless $I_{\rm D}$ we can derive Eq. (\ref{26}) in two distinct ways. If we demand local invariance of the action under $\psi\rightarrow e^{i\alpha(x)}\psi$ and $\psi\rightarrow e^{i\gamma^5\beta(x)}\psi$, we can introduce $A_{\mu}$ and $S_{\mu}$ by minimal coupling or by changing the geometry. The two potentials needed for electromagnetism can thus be put on a completely equal footing. Now a drawback in using  a $B_{\mu}$-dependent $W^{\lambda}_{\phantom{\alpha}\mu\nu}$  is that with it parallel transport is path dependent, with the geometry being a Weyl geometry rather than a Riemannian one. However, with $iA_{\mu}$ the geometry associated with $I_{\rm EFF}$ is a regular Riemannian one that uses only the connections given in Eq. (\ref{17}). Thus by using $iA_{\mu}$ instead of $B_{\mu}$  we convert a Weyl geometry into a Riemannian one.

\section{The Nature of Torsion}

While we have seen that  the axial 4-vector $S_{\mu}$ gives electromagnetism a chiral structure, we need to comment on the fact that experimentally there is no apparent sign of $S_{\mu}$. Moreover, since $S_{\mu}$ is associated with torsion it is not simply a typical spacetime axial vector field. To underscore the special nature of torsion, we note that even if the standard (torsionless) Riemann tensor is zero, torsion is not obliged to vanish. Torsion could thus exist in a spacetime with no Riemann curvature at all. In a space that is flat as far as the geometry of its four spacetime $x^{\mu}$ coordinates is concerned, we note that since the Minkowski metric is independent of the $x^{\mu}$,  a non-zero torsion might not depend on the $x^{\mu}$ coordinates either. Given the antisymmetry of $Q_{\lambda\mu\nu}$, we can thus envisage that $S_{\mu}$, and thus concomitantly its monopole source $K^{\mu}$ as well,  might depend instead on a set of Grassmann coordinates, coordinates that anticommute with each other. 

To realize this possibility, on comparing Eqs. (\ref{5}) and (\ref{6}) with Eq. (\ref{27}), we can consider the possibility that the torsion can  written as $Q_{\lambda\mu\nu}=\nabla_{\lambda}A_{\mu\nu}$, where $A_{\mu\nu}$ is an antisymmetric rank two tensor. In \cite{footnote7} it was suggested that this $A_{\mu\nu}$ could be the antisymmetric part of a 16-component metric tensor. To this end, we now note that if we introduce a set of Grassmann vierbeins $\xi^a_{\mu}$, then the quantity $A_{\mu\nu}=\eta_{ab}\xi^a_{\mu}\xi^b_{\nu}$  will be antisymmetric since $\xi^a_{\mu}\xi^b_{\nu}+\xi^b_{\mu}\xi^a_{\nu}=0$. Thus we can envisage spacetime being enlarged to encompass both ordinary coordinates and Grassmann coordinates, and spaces of this type were constructed in e.g. \cite{Mannheim1984}, where it was shown that a canonical quantization in which vanishing anticommutators  were replaced by non-vanishing ones led to the Dirac equation. As also noted in \cite{Mannheim1984}, because of the Pauli principle finite degree of freedom Grassmann coordinates $\xi^b_{\mu}$ (as opposed to infinite degree of freedom Grassmann fields $\psi(x)$) could not be macroscopically occupied. Consequently, a Grassmann torsion could only be microscopic, with only the sector of electromagnetism that is based on $A_{\mu}$ ordinarily being observable in macroscopic systems. 

Now we had found in Eq. (\ref{28}) that at the classical level the $A_{\mu}$ and $S_{\mu}$ sectors were decoupled from each other. However, according to the Dirac action given in Eq. (\ref{26}) both sectors couple to the fermions. Thus quantum mechanically one could have transitions between the two sectors mediated by fermion loops with both  vector and axial vector insertions (axial analog of light on light scattering).  This would be a small effect, and would also be microscopic, with a quantized Grassmann torsion not making any substantial modifications to QED. Thus torsion, and equally magnetic monopoles, might only be manifest microscopically, where they could potentially contribute to physics beyond the standard model \cite{footnote9}. Finally, if the torsion/monopole sector is only manifest microscopically, then macroscopically we can set $\bf{E}^{\prime}$, $\bf{B}^{\prime}$, $\rho_m$, and $\mathbf{J}_m$ to zero, with the standard sourceless Faraday Law then holding for macroscopic electrodynamics.


\begin{thebibliography}{9}
\bibitem{footnote4} This of course does not preclude their possible existence.
\bibitem{footnote9} As an alternative to Grassmann coordinates one could construct the antisymmetric $A_{\mu\nu}$ via $A_{\mu\nu}=C_{ab}V^a_{\mu}V^b_{\nu}$, where $C_{ab}$ is the charge conjugation matrix. Torsion would then be macroscopic, but could still have escaped detection if the bridge between the $A_{\mu}$ and $S_{\mu}$ sectors is purely quantum-mechanical. Another possibility is for $S_{\mu}$ to acquire a large Higgs mechanism-type mass.
\end{thebibliography}
\end{document}